\documentclass[twocolumn,times,trackchanges, dvipsnames]{aastex631}
\usepackage[utf8]{inputenc}
\usepackage{amsmath}

\begin{document}
\title{The FUV Spectrum of FU Ori South}

\author{Adolfo S. Carvalho}
\affiliation{Department of Astronomy; California Institute of Technology; Pasadena, CA 91125, USA}
\author{Gregory J. Herczeg}
\affiliation{Kavli Institute for Astronomy and Astrophysics, Peking University, Beijing 100871, People's Republic of China}
\affiliation{Department of Astronomy, Peking University, Beijing 100871, People's Republic of China}
\author{Kevin France}
\affiliation{Laboratory for Atmospheric and Space Physics, University of Colorado Boulder, Boulder, CO 80303, USA}
\author{Lynne A. Hillenbrand}
\affiliation{Department of Astronomy; California Institute of Technology; Pasadena, CA 91125, USA}

\begin{abstract}
    The eruptive YSO FU Ori is the eponym of its variable class. FU Ori stars are known to undergo outbursts with amplitudes of $>4$ magnitudes in the $V$ band and durations of several decades. Interaction with a binary companion is one proposed outburst trigger, so understanding both components of the FU Ori system is crucial. A recent HST/STIS observation of the FU Ori system clearly resolves its North and South components. We report here on the spectrum of FU Ori South. We detect NUV continuum emission but no FUV continuum, although several bright emission lines consistent with those seen in T Tauri stars are present. The presence of the C II] 2325 multiplet and many H$_2$ lines indicate active accretion. We estimate the extinction to the source and find that the UV spectrum favors $A_V < 4$, contrary to past estimates based on the NIR spectrum. 
\end{abstract}

\section{Introduction}\label{sec:introduction}
A popular proposed trigger of FU Ori outbursts is the interaction with a binary companion during the periastron passage of the pair \citep{bonnell_BinaryOrigin_FUOriObjects_1992ApJ}. The detection of a companion 0$^{\prime\prime}$.5 to the south of FU Ori \citep{Wang_FUOriBinaryDetection_2004ApJ, ReipurthAspin_FUOriBinaryPhotometry_2004ApJ} lent credence to this theory, with \citet{Pueyo_FUOriS_2012ApJ} confirming its physical association through astrometry. Adaptive-optics-assisted integral field spectroscopy with Gemini North/NIFS  \citep{BeckAspin_FUOriBinarySpectrum_2012AJ}
revealed that the near infrared (NIR) spectrum of FU Ori S has atomic 
absorption features like those in low mass young stellar objects (YSOs) with a K5 spectral type. 

The NIR continuum of FU Ori S has a perplexing shape, with the $J$ band indicating significant line-of-sight extinction of $A_V \approx 6$ mag while the $H-K$ band continuum matches $A_V = 0.5$ mag. The Pa$\beta$ and Br$\gamma$ emission gives an accretion rate of $\dot{M} = 3 \times 10^{-8} \ M_\odot$ yr$^{-1}$ \citep{BeckAspin_FUOriBinarySpectrum_2012AJ}, typical of low-mass Class II YSOs \citep{Manara_PPVIIChapter_2023ASPC}. The CO $\Delta \nu = 2$ bandheads in $K$ band are all in emission, as has been seen in some rapidly accreting Class I/II YSOs \citep[e.g.,][]{Doppmann_GVTauEmission_2008ApJ}.

\citet{Pueyo_FUOriS_2012ApJ} combined a low-resolution Palomar/P1640 spectrum of FU Ori S with MIR photometry and SED fitting to find a stellar temperature range of $T_\mathrm{eff} = 4000-5000$ K, consistent with \citet{BeckAspin_FUOriBinarySpectrum_2012AJ} but requiring higher extinction values ($A_V = 8-12$ mag). The conflicting $A_V$ estimates may be due to the exotic sightline to the source, which includes significant contribution from the disk and accretion streamers around FU Ori N \citep{takami_FUOriPolarizedLight_2018ApJ, Perez_FUOriALMA_2020ApJ, Hales_FUOriStreamer_2024ApJ}. 

We present in this Note the first far- and near- ultraviolet (FUV/NUV) spectrum of FU Ori S, which shows emission lines that are consistent with those seen in accreting low mass young stars. We compare the spectrum with the well-known source BP Tau, which is a typical classical T Tauri Star (CTTS) with a similar spectral type and mass accretion rate \citep{Manara_PPVIIChapter_2023ASPC} to FU Ori S.

\section{Data and Calibration} \label{sec:data}
We obtained spectra of the FU Ori system using the Space Telescope Imaging Spectrograph (STIS) as part of HST Guest Observer program 17176 (PI: L. Hillenbrand). The observation was taken using the 52$^{\prime\prime}$X2$^{\prime\prime}$ slit in the G140L (FUV-MAMA), G230L (NUV-MAMA), and G430L (CCD) grating settings. The spectra are available on the Mikulski Archive for Space Telescopes (MAST) and can be accessed via \dataset[doi: 10.17909/3p42-jw31]{https://doi.org/10.17909/3p42-jw31}. We only see the South component in the two bluest spectra. 

In the FUV and NUV spectra, the STIS position angle (PA) is $-131^\circ$ and the 2$^{\prime\prime}$ slit captures the entire system. The separation of $\sim 0.2^{\prime\prime}$ between the North and South traces in the $\mathtt{x2d}$ spectrum is consistent with the $0.48^{\prime\prime}$ and $163^\circ$ binary separation and PA \citep{Pueyo_FUOriS_2012ApJ} projected to the slit PA. The traces are clearly separated and can be extracted individually. The automatic extraction in MAST included only the primary, so we performed aperture extraction of the secondary. We flux-calibrated using a sensitivity function obtained from previous observations. The final spectrum of FU Ori S covers 1100-3200 \AA\ with $R \equiv \lambda/\Delta\lambda = 600$. Between 1700 \AA\ and 2200 \AA\ the observations are consistent with noise. The spatial offset of FU Ori S produces a $\approx 5$ \AA\ blueshift in the wavelength solution. 

The spectrum of BP Tau was obtained in 2002 in the HST GO program 9081 \citep[PI: N. Calvet;][]{Kravtsova_BPTauUVSpectrum_2003AstL} and was accessed through MAST.

\section{The UV Spectrum of FU Ori S} \label{sec:UVSpec}

\begin{figure*}
    \centering
    \includegraphics[width=0.95\linewidth]{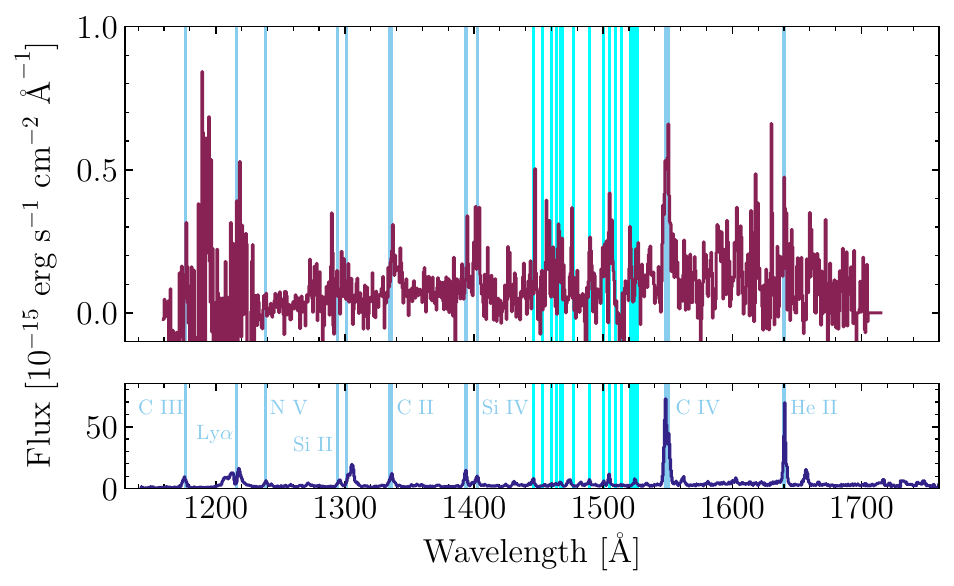}
    \includegraphics[width=0.95\linewidth]{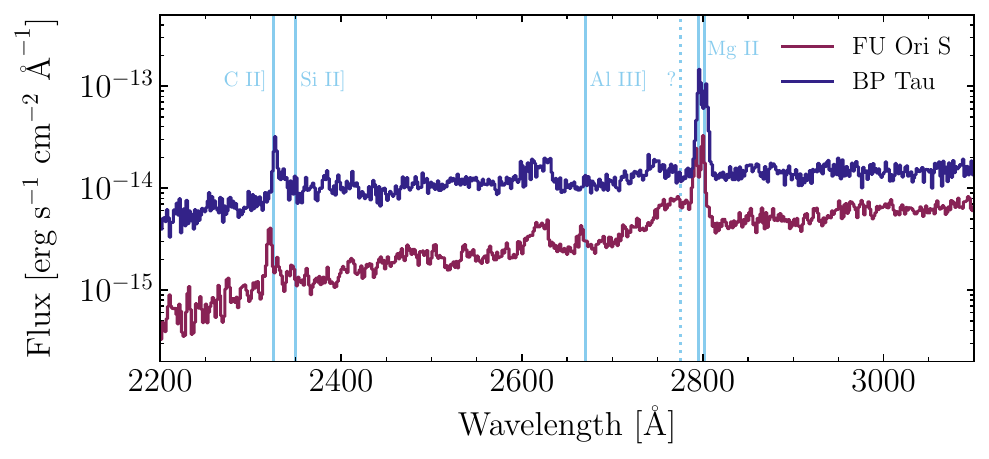}
    \caption{The UV spectrum of FU Ori S, compared with that of BP Tau. Emission lines commonly seen in the spectra of accreting T Tauri stars are marked in vertical solid light blue lines. The dotted line shows the location of the feature at 2775 \AA. Between 1400 \AA\ and 1550 \AA\, the H$_2$ emission lines are marked by cyan vertical lines, some of which are almost as bright as the \ion{C}{4} doublet at 1548/1550 \AA. }
    \label{fig:spec}
\end{figure*}

The spectrum of FU Ori S is shown in Figure \ref{fig:spec}, alongside that of BP Tau. Although the FUV spectrum is not sensitive enough to detect continuum, there are several emission lines that correspond well to those in BP Tau. In particular, Ly$\alpha$, \ion{C}{2} 1334.5/1335.7, \ion{Si}{4} 1393.8/1402.8, \ion{C}{4} 1548.2/1550.8, and \ion{He}{2} 1640 are all clearly detected. 

The NUV spectrum is higher signal-to-noise and shows both continuum and line emission. The NUV \ion{C}{2}] 2325 multiplet, \ion{Si}{2}] 2350 line, \ion{Al}{3}] 2670 line, and the \ion{Mg}{2} 2796/2803 doublet are all prominent and there is a broad emission feature blue-ward of the \ion{Mg}{2} doublet centered at 2775 \AA\ with a Gaussian $\mathrm{FWHM} = 54$ \AA. 

The presence of the \ion{C}{2}] 2325 feature is an indicator of accretion in FU Ori S \citep{Ingleby_NUVEmission_CII_2013ApJ}, so we sought to use it to estimate the $L_\mathrm{acc}$ in the system. First, we used the $\mathtt{fit \_ continuum}$ function in the $\mathtt{specutils}$ Python package to fit a continuum to the spectrum using two regions: 2100 \AA\ to 2200 \AA\ and 2400 \AA\ to 2450 \AA. We then fit a double Gaussian profile to the continuum-subtracted line. 

We adopt the 404 pc distance to the source \citep{Kounkel_LamOriDist_2018AJ, Roychowdhury_FUOriV883OriDist_2024RNAAS} and use the \citet{Whittet_ExtinctionCurve_2004ApJ} extinction law for our extinction corrections. We report the calculated line luminosities assuming both $A_V = 0$ mag and $A_V = 2.5$ mag, the value used by \citet{BeckAspin_FUOriBinarySpectrum_2012AJ}. We measure a line flux of $1.7 \times 10^{-14}$ erg s$^{-1}$ cm$^{-2}$ ($5.2 \times 10^{-12}$ erg s$^{-1}$ cm$^{-2}$) in the multiplet and a luminosity of $8.4 \times 10^{-5} \ L_\odot$ ($0.025 \ L_\odot$) assuming $A_V = 0$ mag (2.5 mag). With the \citet{Ingleby_NUVEmission_CII_2013ApJ} calibration, $\log_{10}(L_\mathrm{acc}) = 1.1 \log_{10}(L_\mathrm{CII]}) + 2.7$, we estimate $L_\mathrm{acc} = 0.02 \ L_\odot$ ($L_\mathrm{acc} = 9 \ L_\odot$). The luminosity for $A_V = 2.5$ mag is almost 50$\times$ greater than the $L_\mathrm{acc} = 0.2 \ L_\odot$ reported by \citet{BeckAspin_FUOriBinarySpectrum_2012AJ}.

Following the same procedure for the \ion{C}{4} 1548/1550 doublet, we compute an integrated line flux of $3.2 \times 10^{-15}$ erg s$^{-1}$ cm$^{-2}$ ($1.1 \times 10^{-12}$ erg s$^{-1}$ cm$^{-2}$), which is $L_\mathrm{CIV} = 1.6 \times 10^{-5} \ L_\odot$ ($ 0.005 \ L_\odot$), assuming $A_V = 0$ mag (2.5 mag). Using the calibration from \citet{yang_HST_TTS_FUV_Survey_2012ApJ}, we estimate an accretion luminosity of $0.005 \ L_\odot$ ($3.90 \ L_\odot$). For $A_V = 2.5$, the $L_\mathrm{acc}$ is lower than that estimated from the \ion{C}{2} NUV multiplet but still more than an order of magnitude greater the  \citet{BeckAspin_FUOriBinarySpectrum_2012AJ} value.

We detect strong emission from \ion{Mg}{2} 2796/2803 doublet, but it is a less reliable tracer of accretion due to heavy outflow absorption \citep{Ardila_TTSsWithHST_GHRS_2002ApJ} and its presence in non-accreting T Tauri star spectra \citep{Ingleby_NUVEmission_CII_2013ApJ}. 

The uncertainty in the $A_V$ to the source translates to uncertainties in $L_\mathrm{acc}$ spanning several orders of magnitude. One potential means of estimating the $A_V$ to FU Ori S from the UV spectrum, independently of existing values derived from the NIR SED, is to use the spectrum of BP Tau as a template. We vary the extinction correction for FU Ori S, normalizing both spectra to 1 at 3000 \AA\ at each step, to find which extinction correction gives the best match to the UV slope of BP Tau. The best-matching relative extinction correction is $\Delta A_V = 2.0$ mag. Given the extinction of $A_V = 0.5$ of BP Tau \citep{herczeg_survey_2014}, this implies an $A_V = 2.5$ mag for FU Ori S, in good agreement with the value adopted by \citet{BeckAspin_FUOriBinarySpectrum_2012AJ}. However, the measured accretion luminosities for \ion{C}{2}], \ion{C}{4}, and \ion{Mg}{2} for this $A_V$ are much larger than their estimates with the NIR hydrogen lines.


  
\section{Conclusions} \label{sec:conclusions}
The UV spectrum of FU Ori S is similar to that of a typical low mass YSO. Although the FUV continuum is undetected and many of the lines are only marginally detected, they are matched to lines seen in another YSO, BP Tau. The NUV is a very good match to the spectrum of BP Tau, with the exception of a broad emission feature at 2775 \AA. The bright FUV/NUV emission, along with measurements of the flux from the \ion{C}{2}], \ion{C}{4}, and \ion{Mg}{2} features indicate that the target is unlikely to be as heavily extincted as previously reported, or that the extinction law departs drastically from that of \citet{Whittet_ExtinctionCurve_2004ApJ}. This may be possible since the dominant source of extinction to FU Ori S may be the circumstellar material and disk of FU Ori N. 

\section{Acknowledgements}
This research was supported, in part, by grant HST-GO-17176.001-A from STScI.

\bibliography{references}{}
\bibliographystyle{aasjournal}

\end{document}